\documentclass[aps,prl,superscriptaddress,twoside,twocolumn,nofootinbib,10pt,%
showpacs,floatfix]{revtex4-1}

\usepackage{amsmath,amssymb}
\usepackage{graphicx,bm}
\usepackage{slashed}
\usepackage{epstopdf}
\usepackage{ulem} 
\usepackage[usenames]{color}
\usepackage{subfigure}
\usepackage{float}

\allowdisplaybreaks

\begin{document}

\title{$D$ and $D^{\ast}$ meson mixing in spin-isospin correlated cold nuclear matter}

\author{Daiki Suenaga}
\email{suenaga@hken.phys.nagoya-u.ac.jp}
\affiliation{Department of Physics,  Nagoya University, Nagoya, 464-8602, Japan}

\author{Bing-Ran He}
\email{he@hken.phys.nagoya-u.ac.jp}
\affiliation{Department of Physics,  Nagoya University, Nagoya, 464-8602, Japan}

\author{Yong-Liang Ma }
\email{yongliangma@jlu.edu.cn}
\affiliation{Department of Physics, Nagoya University, Nagoya, 464-8602, Japan}
\affiliation{College of Physics, Jilin University, Changchun, 130012, China}

\author{Masayasu Harada}
\email{harada@hken.phys.nagoya-u.ac.jp}
\affiliation{Department of Physics,  Nagoya University, Nagoya, 464-8602, Japan}

\date{\today}

\newcommand\sect[1]{\emph{#1}---}
%%%%%%%%%%%%%%%%%%%%%%%%%%%%%%%
\begin{abstract}
We propose to study the mass spectrum of the heavy-light mesons to probe the structure of the spin-isospin correlation in the nuclear medium.
We point out that the spin-isospin correlation in the nuclear medium generates a mixing among the heavy-light mesons carrying different spins and isospins such as $D^+$, $D^0$, $D^{\ast +}$, and $D^{\ast 0}$ mesons. We use two types of correlations motivated by the skyrmion crystal and the chiral density wave as typical examples to obtain the mass splitting caused by the mixing. Our result shows that the structure of the mixing reflects the pattern of the  correlation, i.e., the remaining symmetry. Furthermore, the magnitude of the mass modification provides information of the strength of the correlation.

\end{abstract}
\maketitle
%%%%%%%%%%%%%%%%%%%%%%%%%%%%%%%%%%%%%%%%%%%%%%%%%%%%%%%

%\section{Introduction}

Studying the dense hadronic medium is one of the interesting subjects for understanding the quantum chromodynamics (QCD) in the low-energy region. It will provide an important clue to describe the equation of state inside neutron stars~\cite{Lattimer:2006xb}, and also  it may give some information on the structure of the chiral symmetry breaking~\cite{Friman:2011zz}.

Heavy-light mesons made of a heavy quark and a light quark are expected to be good probes of the properties of nuclear medium.
Although the medium modifications of the properties of heavy-light mesons have been widely studied~\cite{Dmeosn}, to the best of our knowledge, there is no explicit statement on the mixing as a single state 
among heavy-light mesons carrying different spins, such as the pseudoscalar $D$ and the vector $D^{\ast}$ mesons, in medium in the literature.

In this Brief Report, we shall discuss the mixing between the heavy-light mesons carrying different spins such as  $D$ and $D^\ast$ mesons in the nuclear medium caused by the existence of the spin-isospin correlation which is expected in, e.g., the skyrmion crystal~\cite{PV09}, 
the chiral density wave phase~\cite{CDW}, and so on. In the literature, the density at which the spin-isospin correlation becomes significant depends on the model. In the recent Skyrme model calculation including the vector meson effect~\cite{Ma:2013ooa},
the pion has a $p$-wave condensation whose size becomes on the order of a few 100 MeV at about the normal nuclear density $\rho_0$. 
In the chiral density wave phase, on the other hand, the pion develops the position-dependent vacuum expectation value (VEV) in the high density region above $2.4\rho_0$~\cite{CDW}. This VEV implies the existence of the strong spin-isospin correlation.

We start with a set of heavy-light mesons which makes two doublets of isospin as well as two doublets of heavy-quark spin symmetry such as $D^+$, $D^0$, $D^{\ast +}$, and $D^{\ast 0}$. In the heavy quark limit, the set is characterized by the spin of the light cloud surrounding the heavy quark~\cite{Manohar:2000dt}: When the spin of the light cloud is $J_l$, the set of heavy-light mesons are made of mesons of spin $J_l+1/2$ and $ J_l-1/2 $. Since the set carries isospin $1/2$, it includes $2( 4 J_l + 2 )$ states, which are all degenerated in mass at the heavy quark limit and the isospin limit. For example, $D$ and $D^*$ mesons are specified by $J_l=1/2$, and then the $D$ meson carries spin $0$ and $D^*$ carries spin $1$, so that there are eight states.

Let us consider the mass splitting of the above $ 2(4 J_l + 2) $ states in the nuclear medium when there exists a correlation between the isospin and the spin which causes the position-dependent pion condensation~\cite{Kunihiro}. The heavy-light mesons interacting with the pion pick up the effects of the condensation which generates a mixing to modify the mass spectrum. In the heavy quark limit, two states are related to each other by the  heavy quark symmetry to form a ``heavy-quark pair''. $(4 J_l + 2 )$ pairs, which are degenerated in the vacuum, are all separated in the most general pattern of the spin-isospin correlation. We would like to stress that the pattern of the mass splitting reflects the remaining symmetry (or the type of the correlation) in the medium: When the medium has the diagonal SU(2) subgroup of the light-spin SU(2)$_l$ and the isospin SU(2)$_I$,  $(4 J_l + 2 )$ heavy-quark pairs are split into $2 J_l$ degenerate pairs and $(2J_l+2)$ pairs.

In the following we explicitly show some examples of the pattern of the mass splitting among $D$ and $D^*$ mesons. For this purpose, we take the following simple model describing the pion and heavy-light meson interaction~\cite{Manohar:2000dt}:
\begin{eqnarray}
{\cal L}_{\rm heavy} & = & 
{} - {\rm Tr}\left[ \bar{H}_a (i v \cdot D )_{ba}H_b\right] 
\nonumber\\
& & 
{} + g_\pi {\rm Tr }\left[
  \bar{H}_a H_b\gamma_\mu\gamma_5 \mathbb{\alpha}_{\perp ba}^\mu \right]
\ ,\label{eq:heavyL}
\end{eqnarray}
where $v^\mu$ denotes the velocity of the heavy-light meson. The indices $a$ and $b$ stand for the isospin indices. 
The covariant derivative $D_\mu$ is defined as $D_\mu = \partial_{\mu} - i\alpha_{\parallel \mu}$, where 
$\alpha_{\parallel,\perp \mu} = (\partial_\mu\xi \cdot \xi^\dag \pm \partial_\mu  \xi^\dag \cdot \xi)/2i$ and $\xi = \sqrt{U} = \exp(i\tau^A\,\pi^A/2f_\pi)$ with the Pauli matrices $\tau^A$ ($A = 1,2,3$). 
The field $H$ is the doublet of heavy-light mesons with the expression
\begin{eqnarray}
H & = & \frac{(1+v\hspace{-0.17cm}\slash)}{2}[D^{\ast \, ;
\mu}\gamma_\mu + i D \gamma_5].\label{eq:hgphys}
\end{eqnarray}
In the Lagrangian \eqref{eq:heavyL} the coupling constant $g_\pi$ is determined from the decay $D^\ast \to D \pi$ as $|g_\pi| = 0.56$~\cite{Harada:2012km}.

When the nuclear medium has a correlation between the spin and the isospin, the position dependent pion condensation will occur to generate non-zero condensates corresponding to $\alpha_{\perp\mu}^A$ and/or $\alpha_{\parallel\mu}^A$ [denoted as $\langle \alpha_{\perp \mu}(t,\vec{x})\rangle$ and $\langle \alpha_{\parallel \mu}(t,\vec{x})\rangle$, respectively]. In the equilibrium case, the pion condensation does not depend on the time so that $\langle \alpha_{\perp 0}^A(\vec{x}) \rangle = \langle \alpha_{\parallel 0}^A(\vec{x}) \rangle = 0$. Since the velocity of heavy-light mesons only has a zero component, $\langle \alpha_{\parallel i}^A (\vec{x})\rangle $ does not contribute to the present calculation. We are interested in the heavy-light meson mass spectrum, so we take the spatial components of the residual momentum of the heavy-light meson to be zero. Then the inverse propagator matrix is written as
\begin{eqnarray}
\Delta^{[i,j]}_{ab} & = &  \left( 
 \begin{array}{ccc} 
 \Delta_{ab} &{} - \Delta^j_{ab}\\ 
  \Delta^i_{ab} & \Delta^{ij}_{ab}\\ 
 \end{array}  
\right) ,\label{eq:InvP}
\end{eqnarray}
where $i, j = 1,2,3$ are spatial indices for the $D^{\ast}$ meson and 
\begin{eqnarray}
\Delta_{ab} & = & {}  2k_0  \delta_{ab} , \nonumber\\
\Delta_{ab}^i & = & {}  -i g_\pi \left\langle \alpha_{\perp}^{iA}\right\rangle \left(\tau^A\right)_{ba}, \nonumber\\
\Delta^{ij}_{ab} & = & {} 2k_0 \delta_{ab}\delta^{ij} - ig_\pi \epsilon^{ijk}\left\langle \alpha_{\perp}^{kA}\right\rangle \left(\tau^A\right)_{ba}\ ,
\label{prop comp}
\end{eqnarray}
with $k_0$ being the time component of the residual momentum carried by the heavy-light meson. $\langle \alpha_{\perp}^{iA}\rangle$ is understood as $\frac{1}{V}\int_V d^3x\langle \alpha_{\perp}^{iA}(\vec{x}) \rangle$ with $V$ being the matter volume. From this inverse propagator matrix one sees that the $\Delta^i_{ab}$ mixes the $D$ and $D^{\ast}$ meson fields when $\langle \alpha_{\perp}^{ iA} \rangle \neq 0$.

Using the spatial rotation and the isospin rotation, we can always diagonalize $\langle \alpha_{\perp}^{iA} \rangle$ such that it takes nonzero values only for $i=A$. There are several choices for the pattern of the condensation. Here we pick up the following two patterns to show how the mass spectrum of the heavy-light mesons is modified:
\begin{eqnarray}
&& \mbox{(Pattern I)} : \ 
  \langle \alpha_{\perp }^{iA} \rangle = \alpha \delta^{iA}\ ,
\nonumber\\
&& \mbox{(Pattern II)} : \
  \langle \alpha_{\perp}^{iA} \rangle = \alpha \delta^{i3} \delta^{A3}
\ ,
\label{pattern}
\end{eqnarray}
where $\alpha$ is a parameter expressing the strength of the spin-isospin correlation with dimension one. 
Typical examples to induce (Pattern I) and (Pattern II) are the hedgehog ansatz of the skyrmion and the chiral density wave, respectively. It should be noticed that, for (Pattern I), the light-spin SU(2)$_l$ and the isospin SU(2)$_I$ are broken down to the diagonal SU(2) subgroup denoted as SU(2)$_{\rm diag}$ by the existence of the above condensate.
In case of (Pattern II), the condensate is invariant under the U(1)$_l$$\times$U(1)$_I$$\times Z_2$ symmetry transformation, where U(1)$_l$ is a subgroup of the SU(2)$_l$ and U(1)$_I$ is of the  SU(2)$_I$. $Z_2$ is a group associated with the ``charge conjugation'' under which both U(1)$_{l}$ and U(1)$_{I}$ charges are inverted simultaneously.

One can easily obtain the change of the mass spectrum by calculating the eigenvalue equation $\det \Delta = 0$ which, for the above two patterns, is expressed as
\begin{eqnarray}
&& \mbox{(Pattern I)} : \ 
\det \Delta  =  \left(2k_0 - g_\pi \alpha\right)^6 \left(2k_0 + 3 g_\pi \alpha\right)^2 
\ ,
\nonumber\\
&& \mbox{(Pattern II)} : \
\det \Delta  =  \left(2k_0 - g_\pi \alpha\right)^4
\left(2k_0 + g_\pi \alpha\right)^4
\ .
\label{eq:patternlimit}
\end{eqnarray}
These results imply that the eight states, which are degenerated in the vacuum, 
are split into six states and two states for (Pattern I), two types of four states for (Pattern II).
These patterns reflect the remaining symmetry in the following way:
Since we are working in the heavy quark limit,  eight states are classified as four heavy-quark pairs. For (Pattern I), these four pairs are split into three pairs making a triplet of SU(2)$_{\rm diag}$ and one pair making a singlet. In the case of (Pattern II), on the other hand, four heavy-quark pairs are split into the following two types: Two heavy-quark pairs carry the same charges under the U(1)$_l$$\times$U(1)$_I$ subgroup: Two heavy-quark pairs with the charges $(+,+)$ and $(-,-)$ are related to each other by $Z_2$ symmetry. While the other two heavy-quark pairs with the charges $(+,-)$ and $(-,+)$ are related by the $Z_2$. We would like to stress that the mass splitting is of the same order as the strength of the correlation parametrized by $\alpha$.

\begin{figure}[htbp]
\includegraphics[scale=0.85]{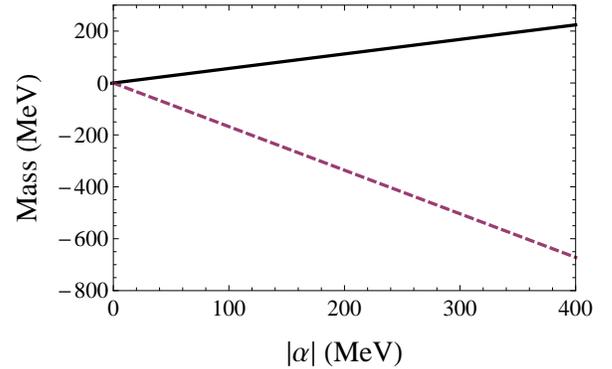}
\caption{(Color online) Dependence of the mass spectrum of $D$ and $D^\ast$ mesons on the strength of the correlation $\alpha$ for (Pattern I). The solid line is the mass of the three heavy-quark pairs making a triplet of SU(2)$_{\rm diag}$, and the dashed line is the one of the one heavy-quark pair making a singlet of SU(2)$_{\rm diag}$.
The vertical axis shows the masses in the medium with the vacuum mass in the heavy quark limit subtracted.
}
 \label{spectrum I}
\end{figure}

\begin{figure}[htbp]
\includegraphics[scale=0.85]{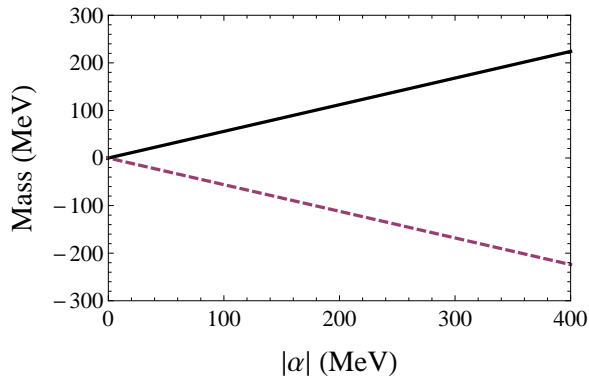}
\caption{(Color online) Dependence of the mass spectrum of $D$ and $D^\ast$ mesons on the strength of the correlation $\alpha$ for (Pattern II). Each of the lines represents the mass of the two heavy-quark pairs. The caption for the vertical axis is the same as that in Fig.~\ref{spectrum I}.}
 \label{spectrum II}
\end{figure}

For expressing the mass spectrum pictorially, we show how it depends on the strength of the correlation $\alpha$ for $g_\pi \alpha >0$
in Figs.~\ref{spectrum I} and \ref{spectrum II}.
From these figures, in addition to the patterns of the mass splitting caused by the spin-isospin correlation, one can easily confirm that the magnitude of the mass modification is of the order of the strength of the correlation parametrized by $\alpha$.

We next consider the effects of the violation of  the heavy quark symmetry
by including the mass difference between $D$ and $D^\ast$ mesons at vacuum.
In this case, the light-spin and the heavy quark spin are not separately conserved but the meson spin, denoted by SU(2)$_J$, is conserved in vacuum. When the spin-isospin correlation is (Pattern I), this SU(2)$_J$ together with the isospin SU(2)$_I$ is broken down to the diagonal SU(2)$_{\rm diag}$. For (Pattern II), on the other hand, the remaining symmetry is U(1)$_J \times$U(1)$_I \times Z_2$, where U(1)$_J$ is a subgroup of SU(2)$_J$ and U(1)$_I$ is a subgroup of SU(2)$_I$. These symmetry structures lead to the following patterns of the mass splitting: For (Pattern I), six $D^\ast$ mesons are split into ${\mathbf 2}$ and ${\mathbf 4}$ representations of SU(2)$_{\rm diag}$, i.e., ${\mathbf 3} \otimes {\mathbf 2} = {\mathbf 2}\oplus{\mathbf 4}$, and two $D$ mesons make ${\mathbf 2}$ which mixes with another ${\mathbf 2}$ from $D^\ast$. 
For (Pattern II), eight states are split into four pairs of $Z_2$ symmetry.
They have the following charges of U(1)$_J \times$U(1)$_I$:
$\left \{ (0,+), (0,- ) \right\}$ from $D$ meson,
$\left \{ (0,+), (0,- ) \right\}$,
$\left \{ (+,+), (-,- ) \right\}$, and $\left \{ (+,-), (-,+ ) \right\}$ from the $D^\ast$ meson
with two  $\left \{ (0,+), (0,- ) \right\}$ states mixing with each other.

\begin{figure}[htbp]
\includegraphics[scale=0.85]{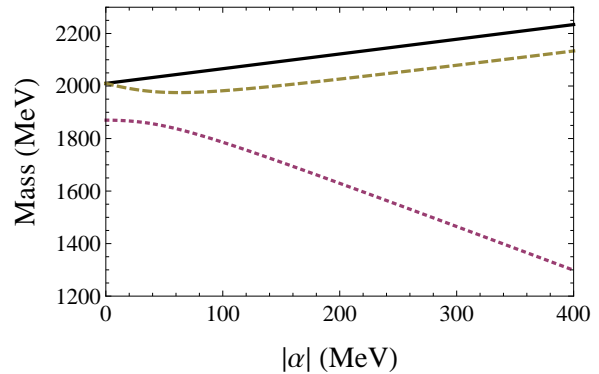}
\caption{(Color online) Dependence of the mass spectrum of $D$ and $D^\ast$ mesons on the strength of the correlation $\alpha$ for (Pattern I). The mass difference of $D$ and $D^{\ast}$ in the vacuum is included. The solid curve represents four degenerate states, and each of the dashed and dotted curves represents two states. }
 \label{spectrum I vio}
\end{figure}

\begin{figure}[htbp]
\includegraphics[scale=0.85]{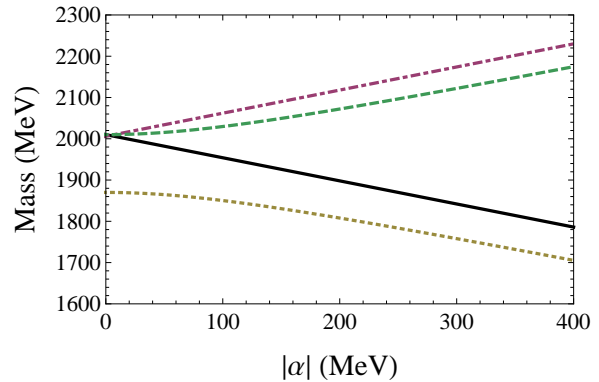}
\caption{(Color online) Dependence of the mass spectrum of $D$ and $D^\ast$ mesons on the strength of the correlation $\alpha$ for (Pattern II). The mass difference of $D$ and $D^{\ast}$ in the vacuum is included. Each of the four curves represents two degenerate states.}
 \label{spectrum II vio}
\end{figure}

We next compute the modification of the mass spectrum by changing $\Delta_{ab}$ and $\Delta_{ab}^{ij}$ in
Eq.~(\ref{prop comp}) as
\begin{eqnarray}
\Delta_{ab} & = & {}  2\left( k_0 - m_D \right)  \delta_{ab} , \nonumber\\
\Delta^{ij}_{ab} & = & {} 2\left( k_0 - m_{D^\ast} \right) \delta_{ab}\delta^{ij} - ig_\pi \epsilon^{ijk}\left\langle \alpha_{\perp}^{kA}\right\rangle \left(\tau^A\right)_{ba}\ ,
\label{prop comp 2}
\end{eqnarray}
while $\Delta^{i}_{ab}$ is the same as that given in Eq.~\eqref{prop comp}.
Using the two patterns shown in Eq,~(\ref{pattern}), we obtain the determinant of the 
inverse propagator as
\begin{eqnarray}
&& \mbox{(Pattern I)} : \ 
\det \Delta  =  \left[2(k_0 - m_{D^\ast}) - g_\pi \alpha\right]^4 \nonumber\\
& &\qquad\qquad\qquad\qquad\quad\;\; {} \times \left[4(k_0 - m_D)(k_0-m_{D^{\ast}}) \right. \nonumber\\
& & \left.\qquad\qquad\qquad\qquad\qquad\;\;\, {} + 4 g_\pi \alpha(k_0 - m_D) - 3 g_\pi^2 \alpha^2\right]^2 
\ ,
\nonumber\\
&& \mbox{(Pattern II)}: \
\det \Delta  = \left[4(k_0 - m_D)(k_0 - m_{D^\ast}) - g_\pi^2\alpha^2\right]^2\nonumber\\
& & {} \qquad\qquad\qquad\qquad\qquad {} \times \left[4(k_0 - m_{D^\ast})^2 - g_\pi^2\alpha^2\right]^2 
\ .
\label{eq:massvio}
\end{eqnarray}
We show the dependence of the  mass spectrum on the strength $\alpha$
in Figs.~\ref{spectrum I vio} and \ref{spectrum II vio}. $D$ and $D^{\ast}$ masses are taken as their empirical values $m_D \simeq 1870$~MeV and $m_{D^\ast} \simeq 2010$~MeV.
In Fig.~\ref{spectrum I vio}, for (Pattern I), the solid curve shows the mass of four degenerated states
which belong to the representation $\mathbf 4$ of the remaining SU(2)$_{\rm diag}$.
Each of the dashed and dotted curves shows the mass for the doublet of the SU(2)$_{\rm diag}$.
In the case of (Pattern II), Fig.~\ref{spectrum II vio} shows the existence of four pairs of $Z_2$ symmetry as explained above.

We finally consider the most general case in which the condensate takes the following form:
\begin{eqnarray}
&& \mbox{(Pattern III)} : \ 
  \langle \alpha_{\perp}^{iA} \rangle = \sum_{j=1}^{3} \alpha_j \delta^{ij} \delta^{Aj}
\ ,
\label{pattern3}
\end{eqnarray}
with $\alpha_1$, $\alpha_2$, and $\alpha_3$ taking arbitrary values.
In this case, the spin and the isospin are completely broken, then,
all eight states have different masses. In the heavy quark limit, they are split into four heavy-quark pairs, with the mass eigenvalue equation  given as
\begin{eqnarray}
\det \Delta  & = & [ 2k_0 - g_\pi \left( \alpha_1 - \alpha_2 + \alpha_3 \right)]^2 \nonumber\\
& & \times [ 2k_0 - g_\pi \left( \alpha_1 + \alpha_2 - \alpha_3 \right)]^2 \nonumber\\
& & \times [2k_0 + g_\pi \left( \alpha_1 - \alpha_2 - \alpha_3 \right)]^2\nonumber\\
& & \times [2k_0 + g_\pi \left( \alpha_1 + \alpha_2 + \alpha_3 \right)]^2 \ .
\end{eqnarray}
This shows that four heavy-quark pairs have different masses as we explained before.

%\section{Conclusions and discussions}

In this Brief Report, we proposed to study the mass spectrum of heavy-light mesons to probe the structure of the spin-isospin correlation in the nuclear medium, which is expected to occur in the skyrmion crystal or the chiral density wave phase.
Our result shows that the structure of the mixing among $D$ and $D^\ast$ mesons reflects the pattern of the  correlation, i.e., the remaining symmetry.
Furthermore, the magnitude of the mass modification provides information of the strength of the correlation. 

Several comments are in order:
\begin{itemize}

\item[(i)] In the general case of the heavy-light mesons of spin $J_l + 1/2$ and $J_l-1/2$
for (Pattern II), $2(4 J_l + 2)$ states are split into $2J_l +1$ types, each of which consists of four states reflecting the $Z_2 \times $SU(2)$_h$ symmetry in the heavy quark limit. When the heavy quark symmetry is violated, the quartet are further split into two doublet of $Z_2$ symmetry.

\item[(ii)] To confront our proposal here with the dense matter properties, a concrete model calculation which explicitly yields the density dependence of $\alpha$ should be performed. This exploration will be done in future publications.

\item[(iii)] The $D$-$D^{\ast}$ mixing discussed in this Brief Report is caused by the $p$-wave pion condensation. A possible environment to realize an enhanced $p$-wave pion condensation is the half-skyrmion phase in the skyrmion matter in which the nuclear tensor force is increasing with density~\cite{Lee:2010sw}.

\item[(iv)] The exploration performed so far is focused on the charmed mesons. Because of the light antiquark in the charmed meson, there might be absorption processes like the antikaon in nuclear matter~\cite{Friedman:2007zza}. However, the present exploration can be extended to anticharmed meson in which the patterns of the mass splitting is the same as that of the charmed meson but the $\alpha$ dependence is flipped. Some explicit calculations will be reported elsewhere.

\item[(v)] The structures of the mass splitting caused by the spin-isospin correlation in the nuclear matter discussed above for the charmed mesons are the same as that for the bottomed mesons. 

\item[(vi)] In this Brief Report, we consider the mixing of the mesons in a heavy-quark doublet.
More generally, we expect mixings among mesons in different heavy-quark multiplets
such as a mixing between $D_1$ ($J^P = 1^+$) and $D^\ast$. 
In addition to the spin-isospin correlation of the nuclear matter, the exploration of the $D^{\ast}$-$D_1$ mixing can also help us to probe the chiral symmetry restoration in the nuclear medium when these two mesons are chiral partners to each other~\cite{heavy-partner}.
This mixing will be studied by extending the present analysis.

\end{itemize}

%%%%%%%%%%%%%%%%%%%%%%%%%%%%%%%%%%%%%%%%%%%%%%%%%%%%%%
%\acknowledgments

We are grateful to M. Rho and S. Yasui for valuable comments and a critical reading of the manuscript. The work of Y.-L.M. and M.H. was supported in part by Grant-in-Aid for Scientific Research
on Innovative Areas (No. 2104) ``Quest on New Hadrons with Variety of Flavors'' from MEXT. 
Y.-L.M. was supported in part by the National Science Foundation of China (NSFC) under 
Grant No.~10905060. 
The work of M.H. was partially supported by 
the JSPS Grant-in-Aid for Scientific Research
(S) No.~22224003, (c) No.~24540266.
 
%%%%%%%%%%%%%%%%%%%%%%%%%%%%%%%%%%%%%%%%%%%%%%%%%%%%%%

\end {document}